\documentstyle[11pt,aaspp]{article}
\newcommand{\postscript}[2]
 {\setlength{\epsfxsize}{#2\hsize}
  \centerline{\epsfbox{#1}}}
\def\ref#1{\par\noindent \hangindent=0.4in \hangafter=1 #1 \par}
\def\eqalign#1{\null\,\vcenter{\openup\jot \m@th
  \ialign{\strut\hfill$\displaystyle{##}$&$
     \displaystyle{{}##}$\hfill \crcr#1\crcr}}\,}
\def\tempest%
{\begin{array}{ccc}
1 & 1 & 1 \\
1 & 1 & 1 \\
4 & 3 & 8
\end{array}}

\begin{document}

\title{The Effects of Amplification Bias in}
\title{Gravitational Microlensing Experiments${}^{1}$}
 
\author{Cheongho Han}
\affil{Ohio State University, Department of Astronomy, Columbus, OH 43210}
\affil{e-mail: cheongho@payne.mps.ohio-state.edu}
\footnotetext[1]{submitted to {\it Astrophysical Journal}, Preprint:
OSU-TA-18/96}
 
\begin{abstract}
 
Although a source star is fainter than the detection limit imposed 
by crowding, it is still possible 
to detect an event if the star is located in the seeing disk of a 
bright star is and gravitationally amplified: amplification bias.
Using a well-constrained luminosity function, 
I show that $\sim 40\%$ of events detected toward 
the Galactic bulge are affected by amplification bias and 
the optical depth might be overestimated by a factor $\sim 1.7$.  
In addition, I show that 
if one takes amplification bias into consideration,
the observed time scale distribution matches significantly better, 
especially in the short time-scale region,
with the distribution expected from a mass-spectrum model in which  
lenses are composed of the known stellar population
plus an additional population of brown dwarfs than it is without 
the effect of the amplification bias.

\end{abstract}
 
 
\bigskip
\bigskip
\bigskip

\newpage
\section{Introduction}

Many candidates of Massive Astrophysical Compact Halo Objects (MACHOs)
are being detected by the  
MACHO (Alcock et al.\ 1993, 1996a), EROS (Aubourg et al.\ 1993, 1995),
OGLE (Udalski et al.\ 1992, 1994), and DUO (Alard 1996) groups.
To maximize the event rate, the searches are being carried
out toward very dense star fields, e.g., the Galactic bulge and
Magellanic Clouds, in which the detection limit is set by crowding.
However, it is possible to detect a lensing event with a source 
star that is below the detection limit, and thus unresolved, provided that  
the star is located in the seeing disk of a bright star and is 
gravitationally amplified. 
This effect is known as ``amplification bias''
(Blanford \& Narayan 1992; Narayan \& Wallington 1994). 
I will call events of this type and the corresponding source 
stars below the detection limit ``faint events'' and ``faint stars'', 
respectively.
Current lensing experiments are adopting an observational
strategy in which they construct templates of resolved stars
to allow fast photometric comparison of source star luminosities.
It may then appear as if faint events cannot be detected because
they are not registered in the template (Bouquet 1993).
However, in the very dense field in which nearly all stars are
blended, the amplified flux of a faint star
will increase the flux of a nearby bright star, and thus 
an event seemingly with the bright star as a source star can be detected.
I will call the stars registered in the template image 
``bright stars'' as opposed to ``faint stars''.
Therefore, the current lensing searches suffer from 
amplification bias (Nemiroff 1994).

In this paper, using a well-constrained luminosity function (LF)  
I show that $\sim 40\%$ of events detected toward
the Galactic bulge are affected by amplification bias and that
the optical depth might be overestimated by a factor $\sim 1.7$. 
In addition,  I show that by taking amplification bias into consideration,
the observed time scale distribution matches 
significantly better, especially in the short time-scale region, 
with the distribution expected from a mass spectrum model 
in which the lenses are composed of the known stellar population 
plus an additional population of brown dwarfs than it does 
without taking account of amplification bias.

\vfill\eject

\section{Faint Lensing Events}

For the detection of a faint event with a faint star flux
$L<L_{DL}$, the amplification should satisfy the condition
$$
{F\over F_0} = 
{L_b + A_{\rm min}L \over L_b + L}
={3\over \sqrt{5}},
\eqno(2.1)
$$
resulting in the minimum required amplification of 
$$
A_{\rm min} = {(3/\sqrt{5})(L_b+L)-L_b
\over L_{\rm }}.
\eqno(2.2)
$$
Here $L_b$ and $L_{DL}$ are the fluxes of the blended bright star and the 
detection limit. 
The factor $3/\sqrt{5}$ in above equations
is included so that an event can be detected when it is amplified 
more than $A=3/\sqrt{5}$ of the combined flux of the bright and faint stars.
The amplification is related to the lensing parameters by
$$
A(u) =
{
u^2 + 2 \over
u(u^2+4)^{1/2}
};\qquad
u^2 = \beta^2 + {(t-t_0)^2\over t_{\rm e}^2},
\eqno(2.3)
$$
where $u$ is the lens-source separation in units of the 
Einstein ring radius $r_{\rm e}$,
$\beta$ is the impact parameter, $t_0$ is the time
of maximum amplification.
The Einstein ring radius is related to the physical parameters of a lens by
$$
r_{\rm e} = \left( 
{ 4GM_L \over c^2 }
{ D_{\rm ol} D_{\rm ls} \over D_{\rm os} }
\right)^{1/2},
\eqno(2.4)
$$
where $M_L$ is the mass of the lens, and $D_{\rm ol}$, $D_{\rm ls}$, and 
$D_{\rm os}$ are the distances between the observer, lens, and source.
The Einstein time scale is related to $r_{\rm e}$ by
$t_{\rm e} = r_{\rm e}/v$, where $v$ is the transverse speed of the lens 
relative to the observer-source line of sight.
The maximum allowed impact parameter for detection is 
related to the minimum required amplification by
$$
\beta_{\rm max} = 
\left[ 2\left( 1-A_{\rm min}^{-2}\right)^{-1/2} - 2 \right]^{1/2}.
\eqno(2.5)
$$
For faint events, one can detect only the portion of a light curve 
above the threshold $L_{b}$ and the event therefore mimics that 
of a shorter time-scale event.
Then what is measured is not $t_{\rm e}$ but the effective time scale
$t_{\rm eff}$, which is the half of the duration of an event above
the threshold.
The ratio between these two time scales is 
$$
\eta (\beta ) =
{t_{\rm eff} \over t_{\rm e}}
=
\left[ \beta_{\rm max}^2-\beta^2 \right]^{1/2} \le 1.0.
\eqno(2.6)
$$

\section{The Fraction of Amplification-biased Events}

Due to amplification bias, each individual bright star works effectively 
as multiple source stars.
Let ${\cal B}$ the effective number of source stars per single 
bright star: amplification bias factor. 
This factor is computed by 
$$
{\cal B} (L_b) = 
1 + 
\int_{0}^{L_b} dL \Phi (L) \beta_{\rm max} (L_b,L) 
\langle \eta \rangle,
\eqno(3.1)
$$
where $\Phi (L)$ is the LF of Galactic bulge stars normalized to the area of 
the seeing disk around each bright star.
The average seeing of the current experiments is $\sim 2''$.
However, when a faint source is located at the edge of the seeing disk of 
the bright star, one can isolate its magnified image from the bright star.
I therefore set the undistinguishable separation between images 
at  $\Delta\theta = 1''\hskip-2pt.5$, i.e., 
$\Phi (L)$ is normalized for stars in the area $\pi (1''\hskip-2pt.5)^2$.
In equation (3.1) the value $\beta_{\rm max}$ is included because only
events with $\beta < \beta_{\rm max}$ can be detected.
In addition, the factor 
$\langle\eta\rangle =\int_0^{\beta_{\rm max}} \eta (\beta )
d\beta/\beta_{\rm max}$ enters
because the detection efficiency, $\epsilon$, decreases as the
time scale decreases.
For the moment I assume that the efficiency is linearly proportional to
the time scale. 
The adopted values of $\Delta \theta$ and the functional form 
$\epsilon (t_{\rm e})$ is subject to some uncertainties,
so I will discuss other cases in \S\ 4.
Then the total effective number of source stars is 
obtained by integrating ${\cal B}(L_b)$ for all bright source stars 
weighted by the LF;
$$
{\cal B}_{\rm tot} = \int_{L_{DL}}^{\infty} dL_b
\Phi (L_b) {\cal B}(L_b)
$$
$$
=
\int_{L_{DL}}^{\infty} dL_b \Phi (L_b)
+
\int_{L_{DL}}^{\infty} dL_b \Phi (L_b) 
\int_{0}^{L_b} dL \Phi (L) \beta_{\rm max} (L_b,L) 
\langle\eta\rangle .
\eqno(3.2)
$$
The first and second terms in equation (3.2) are 
the number of bright stars in the seeing disk and the 
additional faint stars that effectively work as source stars due to 
the amplification bias effect, respectively.

The model LF is constructed as follows.
For stars brighter than the de-reddened $I$-band mag
of $I_0 = 18.2$, I adopt the LF
determined by J.\ Frogel (1996, private communication),
and in the range $18.2 \le I_0 \le 22.4$,  
I use the LF determined by Light et al.\ (1996). 
To extend the LF beyond even this limit, I adopt 
the LF of stars in the solar neighborhood determined by
Gould, Bahcall, \& Flynn (1996).
I address below the uncertainty caused by the 
difference in the bulge and disk stellar populations.
I adopt a distance $R_0=8.0\ {\rm kpc}$ to the Galactic bulge stars.
The model LF is presented in Figure 1 in units of ${\rm stars}\ 
{\rm mag}^{-1}\ {\rm arcmin}^{-2}$.
Current experiments reach the detection limit 
when the stellar number density of Galactic bulge fields arrives at
$\sim 10^6\ {\rm stars}\ {\rm deg}^{-2}$
(C.\ Alcock 1996, private communication).
Based on the model LF this number density corresponds to  
$I_0 = 18.2\ {\rm mag}$ or $M_{I}=3.7$.

With the model LF and the corresponding detection limit,
the amplification bias factor ${\cal B}$ is computed by equation (3.1) 
along with equations (2.2), (2.5), and (2.6). 
\footnote{
An additional threshold for detection might be imposed 
by the flux from other stars located in 
the seeing disk, background flux.
However, flux from stars much fainter than $L_b$ will spread smoothly 
over the seeing disk and it will be subtacted during image process.
Some stars just below $L_b$ that are located in the seeing disk will 
cause the threshold to be higher.
On the other hand, stars that are near but not inside the seeing disk will 
make the bright star appear fainter and so cause a lower threshold.
To lowest order, these two effects cancel one another.
}
The resultant values of ${\cal B}$ are shown as a function of 
bright star mag $I_b$ in Figure 2.
Because the LF increases as 
$L_b$ decreases, combined with the fact that the most probable 
faint events have unamplified faint-star flux 
just below $L_b$, the value ${\cal B}$ increases as $L_b$ decreases.
Since the event rate is directly proportional to the number of source 
stars, the average increase in event rate due to the amplification 
bias effect is determined by
$$
\langle{\cal B}\rangle = 
{{\cal B}_{\rm tot}
\over 
\int_{L_{DL}}^{\infty} dL_b \Phi (L_b)},
\eqno(3.3)
$$ 
resulting in $\langle{\cal B}\rangle = 1.65$.
That is, on average each source star works effectively as $\sim 1.7$ stars,
and thus the event rate increases by the same factor.

The distribution of faint source-star brightness 
for a given template star with a flux $L_b$ is obtained by
taking the derivative of ${\cal B}$, i.e., 
$$
{d{\cal B} (L_b,L)\over dL} = 
\delta(L-L_b) +
\beta_{\rm max} (L_b,L)
\langle \eta \rangle \Phi (L).
\eqno(3.4)
$$
Note that $\int_{L_{DL}}^{\infty} dL\delta(L-L_b) = 1$.
The distributions $d{\cal B}(L_b,L)$ for bright stars with 
$I_b = 15,\ 16,\ 17,\ 18\ {\rm mag}$ are shown in the left 
panel of Figure 3.
The total event distribution for all bright stars is computed by
$$
d{\cal B}_{\rm tot} (L)= 
\int_{L_{DL}}^{\infty} dL_b \Phi (L_b) d{\cal B}(L_b,L)
$$
$$
= \Phi (L_b)dL_b + \int_{L_{DL}}^{\infty} dL_b \Phi (L_b) 
\left[ \beta_{\rm max} (L_b,L)
\langle \eta \rangle \Phi (L) dL\right],
\eqno(3.5)
$$
where the first and second terms represent the event rate distribution 
without and with the amplification bias effect, respectively.
When the amplification bias effect is not included, the distribution is just
proportional to the LF, i.e., only the first term.
The total event distribution is shown in the right panel of 
Figure 3, in which the contribution to the distribution by faint 
events are shaded.
The faint events comprises $\sim 40\%$ of the total events and the
unamplified mag of faint source stars extends up to $I_0 \sim 23$.
For very faint stars, i.e., $I_0 \gtrsim 23$, to be detected they 
should be highly amplified, and thus they are rare.
The slim chance of high amplification becomes even slimmer because 
of the drop in efficiency.
Therefore, the uncertainty in the model LF in the very faint region 
does not affect the determination of 
$\langle {\cal B} \rangle$.

\section{Effect of the Amplification Bias on the Time Scale Distribution}

Amplification bias has a major influence  
on the Einstein time scale distribution $f(t_{\rm e})$ because 
one measures $t_{\rm eff}$ instead of $t_{\rm e}$.
How and how much is the time scale distribution affected?
To answer this question I construct time scale distributions
for events toward the Galactic bulge with 
and without including the effect of amplification bias  
using a reasonable mass spectrum model of lenses.
For the construction of $f(t_{\rm e})$, it is required to model the 
mass and velocity distributions. 
For the Galactic bulge mass distribution, I adopt a ``revised COBE'' model
that is based on the triaxial COBE model (Dwek et al.\ 1995) except for the
central part of the bulge.
In the inner $\sim 600\ {\rm pc}$ of the bulge, I adopt the centrally 
concentrated axisymmetric Kent model (1992) since the COBE model does not 
match well in this region.
The model is provided in terms of the light density, $\nu$.
For the disk, I adopt a Gould et al.\ (1996) model which
has of the form
$$
\rho (R,z) = \rho_0 \left[ {4\over 5}\ {\rm sech}^2 \left( 
{z\over h_1}\right)
+
{1\over 5} \exp \left( -{z\over h_2} \right) \right]
\exp \left( -{R-R_0\over 3000}\right),
\eqno(4.1)
$$
where $h_1=323\ {\rm pc}$ and $h_2 = 660\ {\rm pc}$, and
$\rho_0 = 0.436\ M_{\odot}{\rm pc}^{-3}$.
Both the disk and bulge MACHO transverse speed is modeled by a Gaussian.
In the model, the velocity distributions of disk MACHOs have means and 
standard deviations 
of $(\bar{v}_y,\sigma_y)=(220, 30)\ {\rm km\ s}^{-1}$ and
$(\bar{v}_z,\sigma_z)=(0,20)\ {\rm km\ s}^{-1}$.
The projected components of the Galactic bulge velocity
dispersion are computed from the tensor virial theorem and
results in $(\bar{v}_{y},\sigma_y)=(0,93.0)\ {\rm km\ s}^{-1}$
and $(\bar{v}_{z},\sigma_z)=(0, 78.6)\ {\rm km\ s}^{-1}$.
Here the projected coordinates $(y,z)$ are set so that
the axes are respectively parallel and normal to the Galactic plane.
For more details of the mass and velocity distributions, see
Han \& Gould (1995, 1996).

There may be dark lenses as well as lenses from known stellar populations.
The mass spectrum, $f (M_L)$, of the stellar population is 
constructed by using the mass-luminosity relation provided by 
equation (5) of Henry \& McCarthy (1993).
The white dwarf component in the mass range 
$0.5\ M_{\odot} \le M_L \le 0.7\ M_{\odot}$ is included by normalizing its
mass spectrum so that there are $\sim 10$ times more white dwarfs 
than the number of turnoff plus giant stars.
Finally, brown dwarfs in the mass range 
$0.07\ M_{\odot}\le M_L \le 0.09\ M_{\odot}$ are included in the mass 
spectrum making up the rest of the total bulge mass of
$M_{\rm bulge} = 2.1\times 10^{10}M_{\odot}$, which is adopted from 
Zhao, Spergel, \& Rich (1995).
Then the total mass of each lens population $i$ in the Galactic bulge is 
determined by
$$
M_{{\rm pop},i} = M_{{\rm BW},i}
\left( {L_{\rm bulge} \over L_{\rm BW}} \right),
\eqno(4.2)
$$
where $L_{\rm bulge} = \int_{\rm bulge}dxdydz\  
\nu (x,y,z) = 1.8\times 10^{10}\ L_{\odot}$ and 
$L_{\rm BW} = \int_{\ell_{\rm BW}} dD_{\rm ol}  \nu (D_{\rm ol})=
2412\ L_{\odot}\ {\rm pc}^{-2}$ is the total amount of light in 
the bulge and the integrated light seen through a unit area 
(${\rm pc}^2$) of the Baade's Window (BW) (Kent 1992; Dwek et al.\ 1995). 
The integrated mass of each population is obtained from the 
mass spectrum model, i.e.,
$M_{{\rm BW},i} =\int dM_{L}f_i(M_L)M_L$.
With the mass spectrum model I find that 
$M_{{\rm BW},i}=1579$, $339$, and $897\ M_{\odot}\ {\rm pc}^{-2}$, and 
the resulting total masses in the bulge are 
$M_{{\rm pop},i}=
1.18\times 10^{10}$, 
$0.25\times 10^{10}$, and
$0.67\times 10^{10}\ M_{\odot}$ for
the stellar, white dwarf, and brown dwarf populations, respectively.

With these models,  
the event rate distribution of bulge-bulge self-lensing and disk-bulge events
for a {\it single} source star is computed by
equations (3.3) and (3.4) of Han \& Gould (1996).
The lens masses are drawn from the model mass function.
However, the COBE bulge model is given in terms of
light density, $\nu$, and thus the conversion from $\nu$ to $n$ is required
for proper normalization.
This number-to-light density ratio is determined by
$$
{n\over \nu} = 
{N_{\rm BW} \over L_{\rm BW}}.
\eqno(4.3)
$$
Here $N_{\rm BW} = \sum_i N_{{\rm BW},i}$ is the total number of 
objects in the unit area of sky toward BW, in which the number of 
individual component is obtained by 
$N_{{\rm BW},i}=\int dM_L\ f_i(M_L)=3441$, $566$, and
$11357\ {\rm objects\ pc}^{-2}$ for
the stellar, white dwarf, and brown dwarf populations, respectively.
Once again the total amount of light in the same area of sky is
$L_{\rm BW}=2412\ L_{\odot}\ {\rm pc}^{-2}$ [see below eq.\ (4.2)].
I find $n/\nu = 6.31$.
Note that although brown dwarfs comprise only $32\%$ of the total mass,
they account for $74\%$ of the total number density.

Once $f (t_{\rm e})$ is obtained, the total event rate as a 
function of time scale not including the amplification bias effect,
$F_{\rm w/o} (t_{\rm e})$, by monitoring {\it all} source stars 
is obtained by multiplying the total number stars,
$N_{\ast}$, and the total amount of observation time, 
$T$, into $f (t_{\rm e})$; 
$$
F_{\rm w/o} (t_{\rm e}) =  f(t_{\rm e}) N_{\ast}T;\qquad
N_{\ast}T = \left( {\pi \over 2\tau } \right) \sum_{j}^{N_{\rm event}}
{t_{{\rm e},j} \over \epsilon(t_{{\rm e},j})},
\eqno(4.4)
$$
where $N_{\rm event}$ is the actually detected number of events.
Note that $N_{\ast}$ is the number of stars in the template, and thus 
only resolved stars.
The MACHO group reported $N_{\rm events} = 39$ events in the 
first year bulge season, resulting
in an optical depth of $\tau = 2.4\times 10^{-6}$ for all types of
stars (Alcock et al.\ 1996b), 
which implies $N_{\ast}T=1.6\times 10^{9}\ {\rm days}$.

On the other hand, the construction of the total time scale distribution 
with amplification bias, $F_{\rm with}(t_{\rm eff})$, requires additional 
processing which is described below.  
For a single bright star, the factor $\eta$ is distributed as
$$
g(\eta ) = \int_{L_{DL}}^{\infty}dL_b \Phi (L_b)
\int_{0}^{L_b} dL \Phi (L) \int_0^{\beta_{\rm max}} 
d\beta \ \delta 
\left[ \eta-\left( \beta_{\rm max}^2 -\beta^2\right)^{1/2}\right]
\left[ \int_{L_{DL}}^{\infty} dL_b \Phi(L_b) \right]^{-1}.
\eqno(4.5)
$$
Once $g(\eta)$ is obtained, the effective time scale distribution 
of faint events $F_{f}(t_{\rm eff})$ is obtained from 
$F_{\rm w/o}(t_{\rm e})$ by
$$
F_{f}(t_{\rm eff}) = 
\int_{0}^{1} d\eta g(\eta ) \int_0^{\infty} dt_{\rm eff} F_{\rm w/o}(t_{\rm e}) 
\delta ( t_{\rm eff} - \eta t_{\rm e}).
\eqno(4.6)
$$
Then, one finds the total (both bright and faint) 
event time scale distribution by
$F_{\rm with}(t_{\rm e}) = F_{\rm w/o}(t_{\rm e}) + F_f(t_{\rm e})$.
With the detection efficiency $\epsilon (t_{\rm e})$ provided by
Alcock et al.\ (1996b), the final time scale distributions 
with and without amplification bias are computed by
$ \Gamma_{\rm w/o} (t_{\rm e})= \epsilon (t_{\rm e}) F_{\rm w/o} (t_{\rm e})$ 
and $ \Gamma_{\rm with}(t_{\rm e}) = 
\epsilon (t_{\rm e}) F_{\rm with} (t_{\rm e})$, and 
they are shown in Figure 4.
In the figure, the distributions are compared
with the observed time scale distribution (shaded histogram) obtained by
the MACHO group (Alcock et al.\ 1996b).

There are two major changes in the time scale distribution by taking 
the amplification bias effect into account.
First, the distribution $\Gamma_{\rm with} (t_{\rm e})$, in general, has a 
higher normalization relative to $\Gamma_{\rm w/o} (t_{\rm e})$ 
due to the increase in number of stars that are effectively monitored. 
The increase factor is 
$\langle {\cal B}\rangle = \int dt_{\rm e} \Gamma_{\rm with} (t_{\rm e})
/  \int dt_{\rm e} \Gamma_{\rm w/o} (t_{\rm e}) = 1.65$, 
which matches well with the value determined in \S\ 3 using the 
approximation $\epsilon (t_{\rm e}) \propto t_{\rm e}$.
Second, $\Gamma_{\rm with} (t_{\rm e})$ is shifted toward shorter 
time scale compared to $\Gamma_{\rm w/o} (t_{\rm e})$ due to 
the additional contribution by faint events which mainly have short 
time scales.
Therefore, a significant fraction of short events ($\sim 10\ {\rm days}$),
which could not be explained by known lens populations, 
might be caused by the amplification bias effect.

However, there are some uncertainties in determining 
the contribution to the event rate by faint events.
One of the uncertainties comes from the maximum size of separation at which 
one can isolate the image of the amplified faint star from that of a
bright star. 
I compute the increase factors in event rate for 
$\Delta \theta=1''\hskip-2pt.0$ and $2''\hskip-2pt.0$ 
and find that $\langle {\cal B} \rangle = 1.35$, and 2.38, respectively.
Additional uncertainty comes from the blending of bright stars with
faint stars.
Due to the blending, some fraction of bright events would fail to 
be detected and the measured time scale would be shorter, resulting lower
normalization and additional shift toward shorter time scale in 
$\Gamma (t_{\rm e})$.
However, because of the dominance of the bright star flux, this effect would 
also be small.

\section{The Effects of Amplification Bias on Optical Depth Determination}

When the amplification bias is not taken into consideration 
the optical depth might be overestimated by a factor 
$\langle {\cal B}\rangle \sim 1.7$ because events
are detected by monitoring 
$N_{\ast ,{\rm eff}}=\langle {\cal B}\rangle N_{\ast}$, while  
$\tau$ is determined with $N_{\ast}$ instead of $N_{\ast ,{\rm eff}}$.
On the other hand, the effects of amplification bias on 
$t_{\rm e}$ and $\epsilon$ do not propagate 
to the optical depth determination.
This is because by measuring $t_{\rm eff}$ instead of $t_{\rm e}$,
the time scale decreases by a factor $\sim \eta$ and for the same reason
the detection efficiency decreases by a similar factor, i.e., 
the same factors cancel each other out and thus there is no net effect.
In addition, most additional events due to amplification bias
are expected to have short $t_{\rm eff}$ in which the efficiency
is well approximated as linear.

\section{Conclusion}

I have shown that amplification bias may have significant effects 
on the time scale distribution and the optical depth determination, and thus 
the correction of the bias is very important.
The true distribution $\Gamma_{\rm w/o}(t_{\rm e})$ can be recovered 
statistically with the known weight factor $d{\cal B}_{\rm tot} (L)$ in the 
reverse way that $\Gamma_{\rm with}$ is obtained from $\Gamma_{\rm w/o}$. 
In addition, for some long $t_{\rm eff}$ events for which
very detailed light curves can be constructed, it will be
possible to recover the individual true $t_{\rm e}$ by the fitting the 
light curves using an additional parameter, the unmagnified flux. 
Another way to detect biased events is finding the shift of the  
centroid of source stars caused by the faint star amplification.
This shift was actually detected by Alard (1996)
despite the moderate quality of the image: photographic plate.
An inspiring development in the lensing experiments is that
the time resolution of observations is improving rapidly.
For example, the alert system allows intensive network
observations of candidate events (PLANET, Albrow et al.\ 1996; 
GMAN, Pratt 1996).
The EROS group (Aubourg et al.\ 1995) has carried out a lensing experiment
toward the Magellanic Clouds with a monitoring frequency of up to
46 times per night.
In this way, it will be possible to obtain detailed light curves,
and so $t_{\rm e}$ for a significant fraction of individual events.

\acknowledgments
I acknowledge precious discussions with A.\ Gould.
This work was supported by the grant AST 94-20746 from the NSF.

\newpage
\postscript{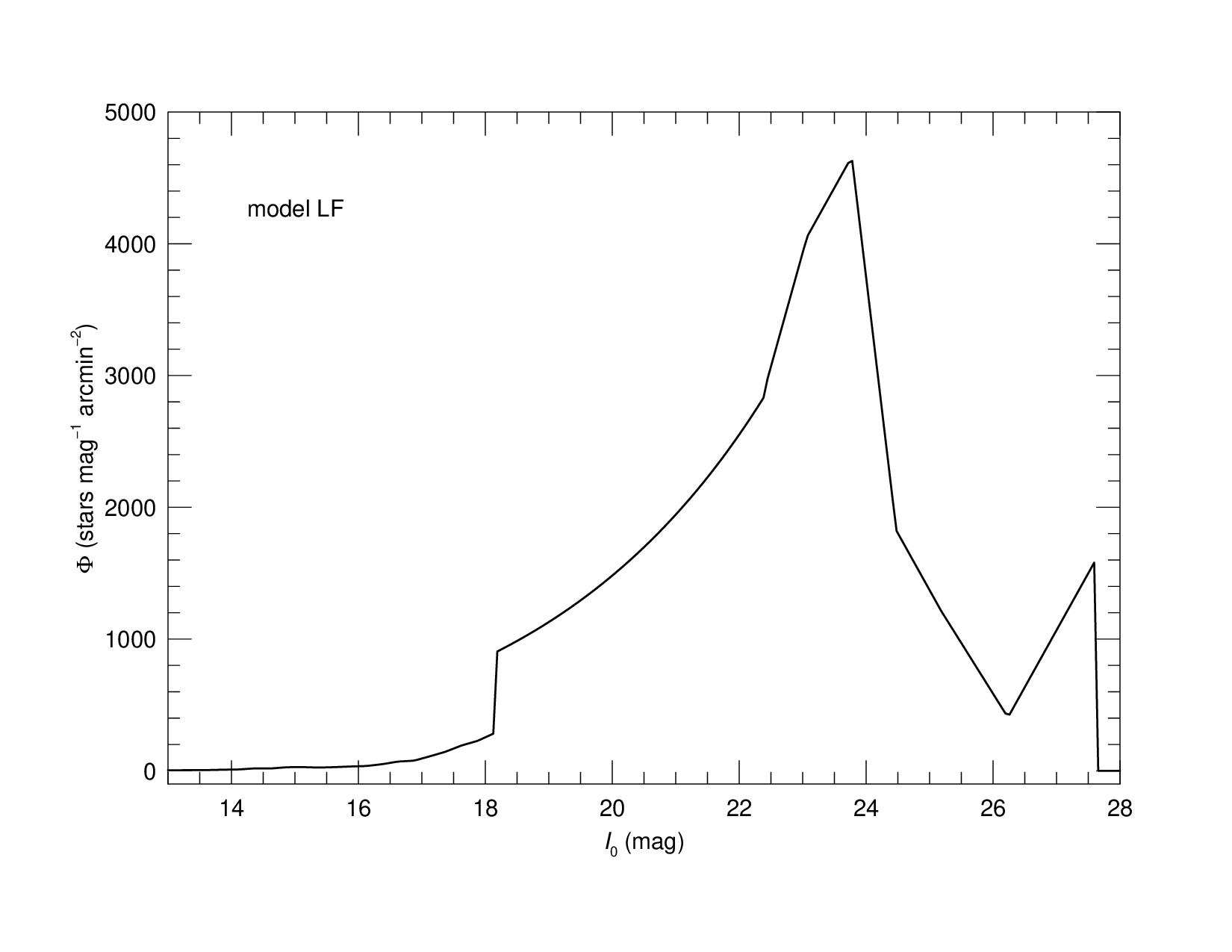}{1.0}
\noindent
{\footnotesize {\bf Figure 1:}\
The model luminosity function in de-reddened $I$-band.
}

\newpage
\postscript{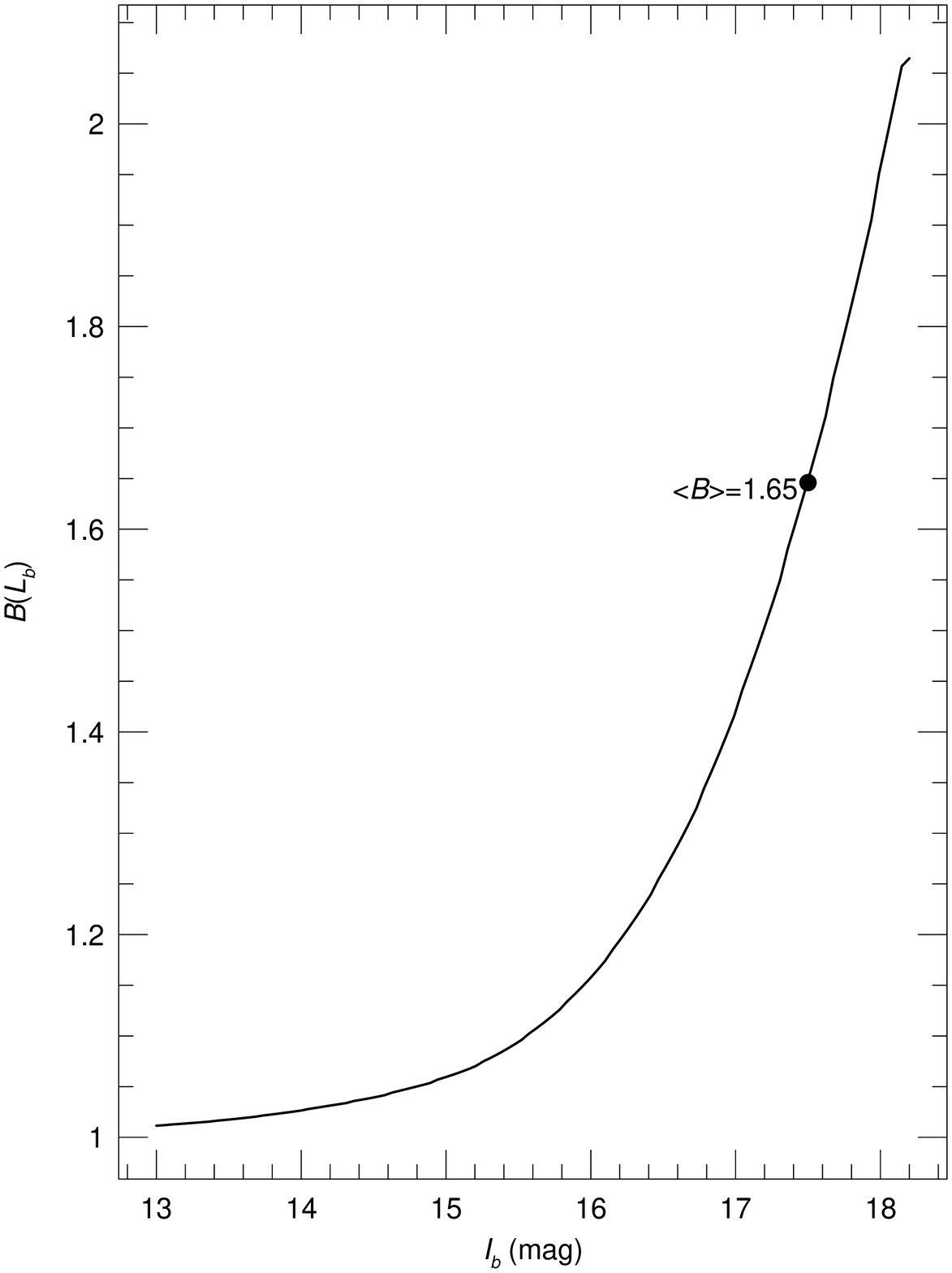}{0.85}
\noindent
{\footnotesize {\bf Figure 2:}\
The amplification bias factors as a function of bright source star mag.
Bacause the LF of bright stars increases, combined with the fact that 
the most probable faint events have unamplified faint-star flux 
just below $L_b$, the value ${\cal B}$ increases as $L_b$ decreases.
}

\newpage
\postscript{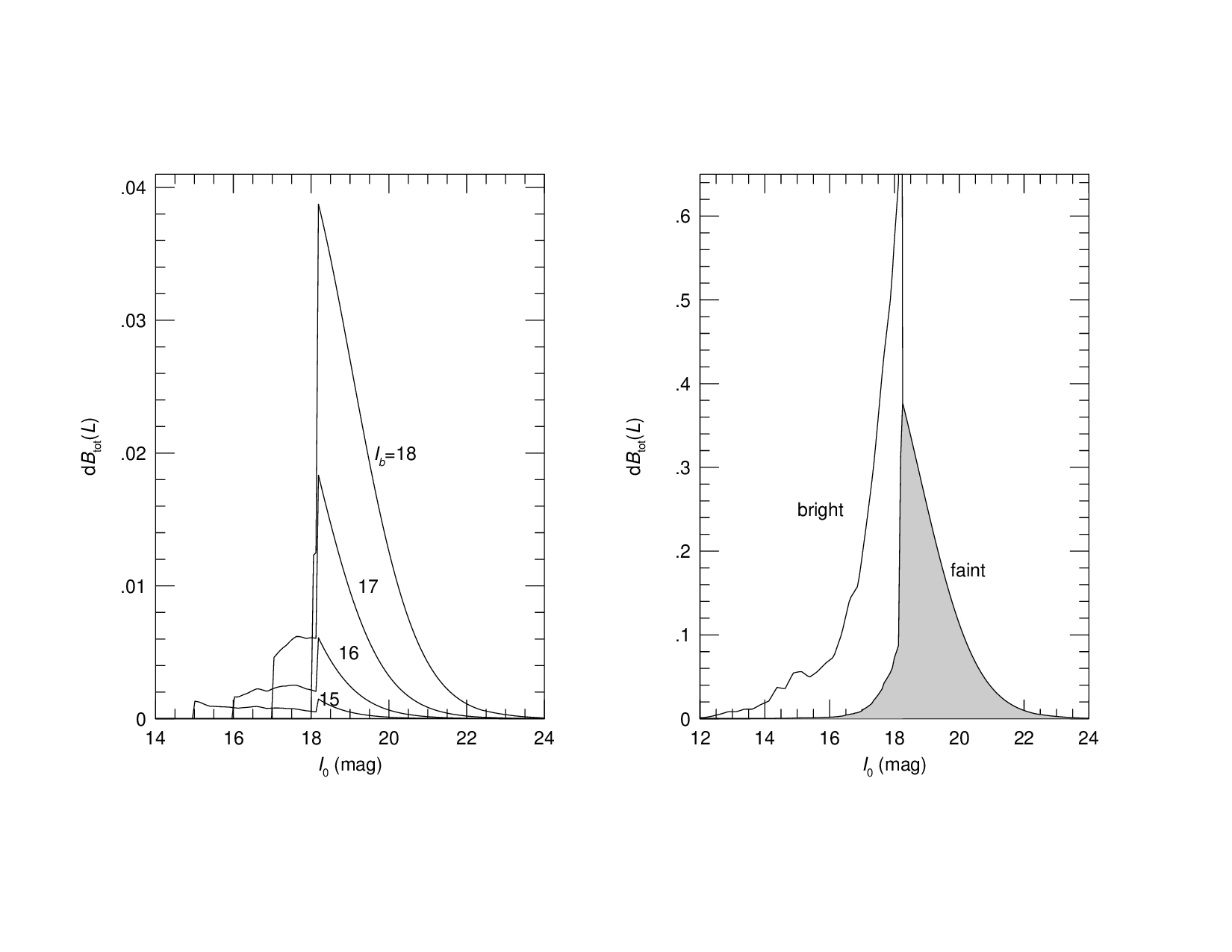}{1.0}
\noindent
{\footnotesize {\bf Figure 3:}\
The distributions of faint source-star brightness for bright star mag
of $I_b = 15$, 16, 17, and 18.
Also shown are the total (both bright and faint) event distributions  
as functions of source star mag.
The contribution by faint events is shaded.
Under the approximation that $\epsilon \propto t_{\rm e}$, faint 
events comprise 40\% of total events.
}


\newpage
\postscript{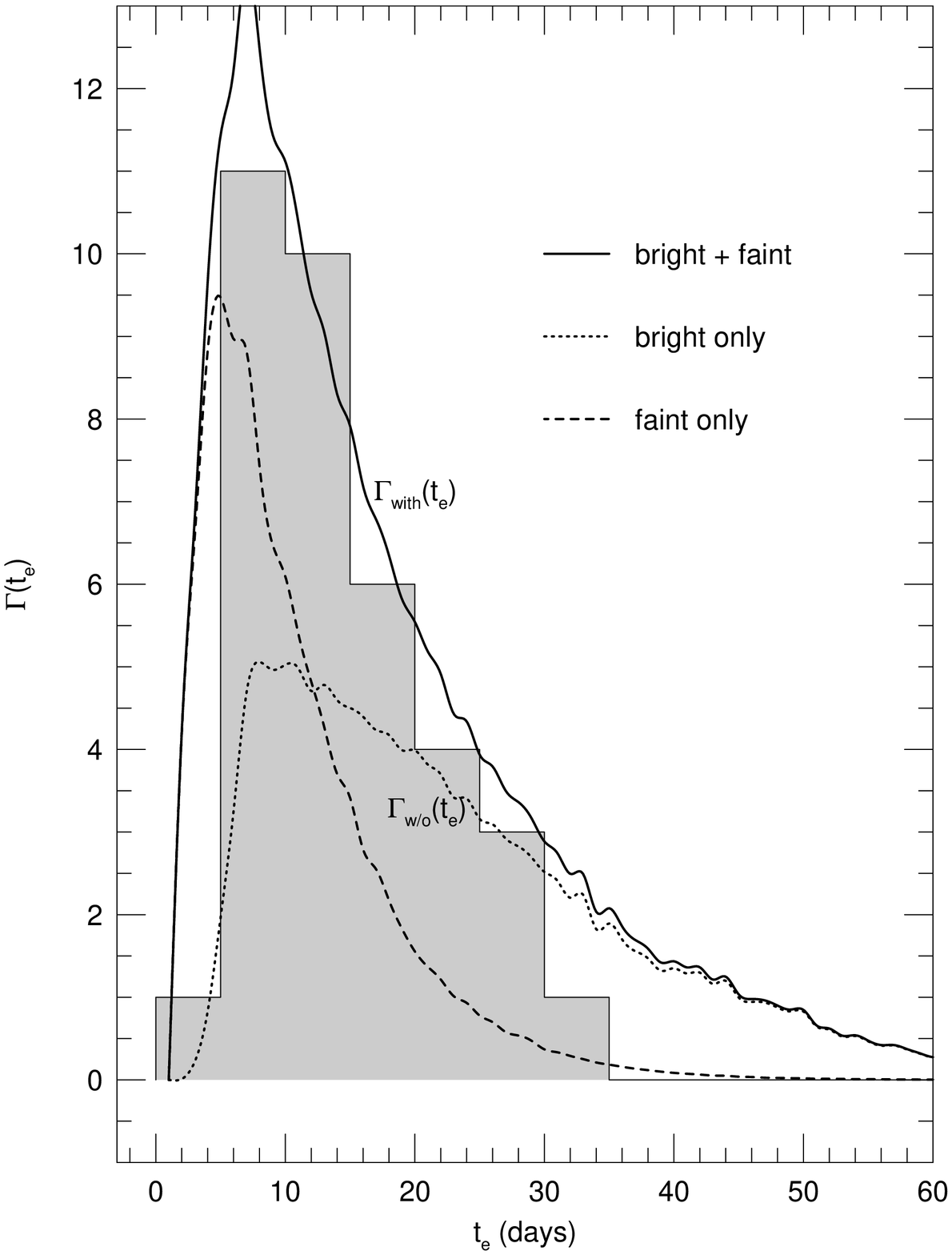}{0.85}
\noindent
{\footnotesize {\bf Figure 4:}\
The time scale distributions for {\it all} stars with and without 
including amplification bias effect, and they are compared with 
the observed distribution obtained by the MACHO group (shaded histogram).
}

\end{document}